\documentclass[%
 aip,
 amsmath,amssymb,
reprint,
]{revtex4-1}

\usepackage{graphicx}
\usepackage{dcolumn}
\usepackage{bm}
\newcommand{\poiss}[2]{\left\{#1,#2\right\}}

\newcommand{\aaa}{{\cal{A}}}
\newcommand{\bbb}{{\cal{B}}}
\newcommand{\ccc}{{\cal{C}}}
\newcommand{\eee}{{\cal{E}}}

\newcommand{\rrr}{{\cal{R}}}
\newcommand{\pq}{\underline{p},\underline{q}}
\newcommand{\qq}{\underline{q}}
\newcommand{\aaaa}{\underline{a}}
\newcommand{\pp}{\underline{p}}
\newcommand{\hh}{\underline{h}}
\newcommand{\ee}{\underline{e}}

\newcommand{\sover}{E_N}
\newcommand{\htilde}{\tilde{h}}
\newcommand{\hhh}{{\cal{H}}}

\begin{document}

\preprint{AIP/123-QED}

\title[Integrable Hamilton\-ians from commuting polynomial families]{
An approach for obtaining integrable Hamilton\-ians from 
Poisson-commuting polynomial families
}

\author{F. Leyvraz}
 \altaffiliation[Also at ]{Centro Internacional de Ciencias.}

\affiliation{ 
Instituto de Ciencias F\'isicas---UNAM, Av.~Universidad s/n, Cuernavaca, 62210, Morelos, M\'exico
}

\date{\today}
\begin{abstract}
We discuss a general approach permitting the identification of a broad class of sets of Poisson-commuting 
Hamilton\-ians, which are integrable in the sense of Liouville. It is shown that all such Hamilton\-ians 
can be solved explicitly by a separation of variables {\em Ansatz}. 
The method leads in particular to a proof
that the so-called ``goldfish'' Hamilton\-ian is maximally superintegrable, and leads to an elementary 
identification of a full set of integrals of motion. The Hamilton\-ians in involution with the ``goldfish'' Hamilton\-ian
are also explicitly integrated. New integrable Hamilton\-ians are identified, among which some have the property of
being isochronous, that is, that all their orbits have the same period. Finally, a peculiar
structure is identified in the Poisson brackets
between the elementary symmetric functions and the set of Hamilton\-ians commuting with the ``goldfish''
Hamilton\-ian: these can be expressed as products between elementary symmetric functions and Hamilton\-ians.
The structure displays an invariance property with respect to one element, and has both a symmetry and
a closure property. The meaning of this structure is not altogether clear to the author, but it turns out to be a 
powerful tool. 
 
\end{abstract}

\pacs{45.05.+x, 45.20.Jj}
\keywords{
integrable systems, polynomials, goldfish equations, Hamiltonian systems
}
\maketitle
%
%
\section{Introduction
}
\label{sec:intro}

In the following, I wish to present a remarkably elementary way of obtaining integrable Hamilton\-ians 
by defining suitable families of Poisson-commuting polynomials. As a motivating example, let us consider 
the following system of ordinary differential equations (ODE's), discussed initially by Calogero in 
[\onlinecite{goldfish}]
and further in the book [\onlinecite{hamil1}]:
\begin{equation}
\ddot{q}_j=2\dot{q}_j{\sum_{k=1(k\neq j)}^N}\frac{\dot{q}_k}{q_j-q_k}.
\label{eq:1}
\end{equation}
This system, known as the ``goldfish''\cite{calorig}, can be solved exactly by the observation that it describes 
the motion of the zeros of a time dependent monic polynomial $S_N(z,t)$
of degree $N$, given by 
\begin{equation}
S_N(z, t)=z^N+\sum_{k=1}^Ns_k(t)z^{N-k},
\label{eq:2}
\end{equation}
such that $\ddot{s}_k(t)=0$. The way to solving (\ref{eq:1}) exactly thus consists in evaluating the initial 
values $s_k(0)$ and $\dot{s}_k(0)$ from $q_k(0)$ and $\dot{q}_k(0)$. One then can readily compute 
the $s_k(t)$ for all times, and the finding of the $q_k(t)$ is thus reduced to the purely {\em algebraic\/}
task of solving for the zeros of $S_N(z,t)$ given the $s_k(t)$. 

There is, however, considerably more structure hidden in this system: first, it can be derived from a Hamilton\-ian.
This follows straightforwardly, as remarked in \cite{hamil}, from the fact that the dynamics of the $s_k(t)$ is 
Hamilton\-ian (since it is free motion). The corresponding motion of the $q_k(t)$ is thus Hamilton\-ian as well, since 
the $q_k(t)$ arise from the $s_k(t)$ by a point transformation. This is obvious for the case of the transformation 
from the $q_k(t)$ to the $s_k(t)$, which is algebraically straightforward. 
This Hamilton\-ian, however, is neither easily described nor analyzed. However, another Hamilton\-ian was stated in 
\cite{periodic}, see also \cite{hamil1}, given by
\begin{equation}
h_1(\pq)=\sum_{r=1}^N e^{p_r}{\prod_{s=1(s\neq r)}^N}\left(
q_r-q_s
\right)^{-1}.
\label{eq:3}
\end{equation}
Here and in the following, underlined quantities such as $\pp$, will always represent an $N$-com\-ponent
vector $(p_1,\ldots,p_N)$. 
That this Hamilton\-ian generates (\ref{eq:1}) as ``Newtonian'' equations of motion is a straightforward
computation. 

Clearly, the solvability of (\ref{eq:1}) implies the integrability of $h_1(\pq)$. Yet several questions remain: 
since the system is
equivalent to free motion, it must be, in fact, maximally superintegrable, that is, there must exist $2N-1$ integrals
of motion of $h_1(\pq)$, of which $N$ can be chosen in involution. Can these be given in a straightforward way?
Can the integration of the Hamilton\-ian (\ref{eq:3}) be realised explicitly? And finally, can this 
be generalised to a broader class of Hamilton\-ians?

Let us first illustrate the method we shall use on this example: one defines the following {\em non-monic\/}
polynomial of degree $N-1$
\begin{equation}
\hhh(z|\pq)=\sum_{k=1}^Nh_k(\pq)z^{N-k}
\label{eq:4}
\end{equation}
by the following conditions:
\begin{equation}
\left.\hhh(z|\pq)\right|_{z=q_k}=e^{p_k}.
\label{eq:5}
\end{equation}
The polynomial is thereby uniquely defined, and the coefficients $h_k(\pq)$ can thus be evaluated. 
It is straightforward to verify that $h_1(\pq)$ is in fact given by (\ref{eq:3}), so that the notation is consistent.

It can now be shown rather simply, using the definition of $\hhh(z|\pq)$ provided by (\ref{eq:5}), that
\begin{equation}
\poiss{\hhh(z|\pq)}{\hhh(w|\pq)}=0
\label{eq:6}
\end{equation}
for all $z$ and $w$, where the Poisson bracket takes its usual form, see (\ref{eq:8}). 
From this follows that all $h_k(\pq)$ are in involution, thus providing a simple proof of the integrability of 
$h_1(\pq)$. $N$ integrals of motion are now explicitly given, and further analysis easily 
yields, as we shall see below, another $N-1$.  We shall also see that this further
provides an approach to solve the Hamilton\-ian (\ref{eq:3}) by separation of variables. While we shall not 
pursue this further here, this might be of interest in the solution of the problem of quantising the
goldfish Hamilton\-ian. While much progress has been made in this direction using Noether
symmetries, see for example \cite{GF-QM1, GF-QM2}, an explicit approach to quantising 
(\ref{eq:3}), allowing for example to display eigenfunctions in configuration space, is still lacking.

As to comparison with earlier works, the following remarks are in order: the most general set of Hamilton\-ians
described in this paper, displayed in (\ref{eq:20}), have been described in various papers within
the context of bi-Hamilton\-ian and quasi-bi-Hamilton\-ian systems\cite{morosi1,blaszak,morosi2,tondo}.
Results concerning their integrability and their amenability to treatment via separation of variables 
are obtained there, but the connection to the quite elementary approach presented here is not apparent 
to the author. 

In the following, we show a general way of obtaining results using an approach
similar to the one sketched above. In Section \ref{sec:general}
we show an easily obtained, yet quite general result. In particular, a family of Hamilton\-ians is displayed in (\ref{eq:20})
which is always integrable. It is further shown how separation of variables may always be achieved under
these circumstances. In Section \ref{sec:translation} we briefly discuss the behaviour of the total momentum, which 
generates the group of translations. In Section \ref{sec:goldfish}, we specify the results to the goldfish Hamilton\-ian
and show a remarkable structure in the Poisson brackets of the elementary symmetric functions with the 
Hamilton\-ians from the set of goldfish Hamilton\-ians. Finally, in Section \ref{sec:conclusions} we present some conclusions.

\section{General approach}
\label{sec:general}
Let us start out with a perfectly general framework of the type sketched in Section \ref{sec:intro}. 
We describe the polynomial $\aaa(z|\pq)$ in the following two ways:
\begin{subequations}
\begin{eqnarray}
\aaa(z|\pq)&=&\sum_{k=1}^Na_k(\pq)z^{N-k},
\label{eq:7a}\\
\aaa(q_k|\pq):=\left.\aaa(z|\pq)\right|_{z=q_k}&=&\alpha_k(\pq).
\label{eq:7b}
\end{eqnarray}
\label{eq:7}
\end{subequations}
Let us here make some notational conventions, to which we shall always adhere. First, the expression
$\aaa(q_k|\pq)$, as well as all similarly constructed expressions, will always be taken in the meaning 
of the definition that immediately follows it in (\ref{eq:7b}). Further,  all polynomials,
which in the following will always be denoted by capital calligraphic letters,
will depend on one or several variables, denoted by $z$, $w$ and so on, and parametrically 
on the variables $\pp$ and $\qq$, which will denote 
vectors of dimension  $N$ of variables $(p_1,\ldots,p_N)$ and $(q_1,\ldots,q_N)$, which obey the usual Poisson 
brackets:
\begin{equation}
\poiss{p_i}{q_j}=\delta_{ij}.
\label{eq:8}
\end{equation}
The polynomials will always be of degree $N-1$, with no restriction on the values of their coefficients.
They are thus uniquely determined by the specification of the values on the $N$ points $q_k$, as
given in (\ref{eq:7b}). These
values will always be given as (lowercase) Greek letters, whereas the coefficients of the polynomial will always be
given as lowercase Latin letters. 

The lone exception to the rule that polynomials are all of degree $N-1$, 
will be the polynomial $\sover(z|\qq)$, defined as
\begin{equation}
\sover(z|\qq)=\prod_{k=1}^N(z-q_k),
\label{eq:9}
\end{equation}
which is, of course, monic of order $N$, and is defined by the fact that it vanishes on all values
$z=q_k$. To underline the difference, we denote this polynomial by a capital Latin letter. 

\subsection{The Poisson bracket formula for polynomials}
\label{subsec:PBpoly}

Let us consider two polynomials, $\aaa(z|\pq)$ and $\bbb(w|\pq)$, where $\bbb(w|\pq)$ is defined
quite analogously to (\ref{eq:7}), by
\begin{subequations}
\begin{eqnarray}
\bbb(w|\pq)&=&\sum_{k=1}^Nb_k(\pq)w^{N-k},
\label{eq:10a}\\
\bbb(q_k|\pq)&=&\beta_k(\pq).
\label{eq:10b}
\end{eqnarray}
\label{eq:10}
\end{subequations}
We now wish to evaluate $\ccc(z,w|\pq)$ defined by
\begin{equation}
\ccc(z,w|\pq)=\sum_{r,s=1}^N\poiss{a_k(\pq)}{b_l(\pq)}z^{N-r}w^{N-s}
\label{eq:11}
\end{equation}
by the simple device of computing its values 
\begin{equation}
\ccc(q_k,q_l|\pq)=\sum_{r,s=1}^N\poiss{a_r(\pq)}{b_s(\pq)}q_k^{N-r}q_l^{N-s}.
\label{eq:12}
\end{equation}

This leads, after some straightforward algebra detailed in Appendix \ref{app:a}, to
\begin{eqnarray}
\ccc(q_k,q_l|\pq)&=&\poiss{\alpha_k(\pq)}{\beta_l(\pq)}\nonumber\\
&+&\left.\frac{\partial\aaa(z|\pq)}{\partial z}\right|_{z=q_k}\frac{\partial\beta_l(\pq)}{\partial p_k}\nonumber\\
&-&\left.\frac{\partial\bbb(w|\pq)}{\partial w}\right|_{w=q_l}\frac{\partial\alpha_k(\pq)}{\partial p_l},
\label{eq:basic}
\end{eqnarray}
which is the basic equation underlying the entire approach. 

The strategy should now be clear: we define polynomials by setting $\alpha_k(\pq)$ and $\beta_k(\pq)$
to values that are, in a sense, simple, so that the Poisson brackets are easily evaluated
 via (\ref{eq:basic}). This yields 
 results which are usually not so trivial for the coefficients $a_k(\pq)$ and $b_k(\pq)$. 

\subsection{The case of one single self-commuting polynomial}

Consider a polynomial $\hhh(\pq)$ defined as usual by 
\begin{subequations}
\begin{eqnarray}
\hhh(z|\pq)&=&\sum_{k=1}^Nh_k(\pq)z^{N-k},
\label{eq:15a}\\
\hhh(q_k|\pq)&=&\overline{\eta}_k(\pq),
\label{eq:15b}
\end{eqnarray}
\label{eq:15}
\end{subequations}
and let us assume that 
\begin{equation}
\poiss{\hhh(z|\pq)}{\hhh(w|\pq)}=0.
\label{eq:16}
\end{equation}
Then the $h_k(\pq)$ are all in involution, and the Hamilton\-ians $h_k(\pq)$ are all integrable in the sense 
of Liouville.

To generate a set of integrable Hamilton\-ians, it is therefore sufficient to give a set of
functions $\overline{\eta}_k(\pq)$ for $1\leq k\leq N$, leading to (\ref{eq:16}) via (\ref{eq:basic}). I 
am not aware of a general solution
to this problem, but the following {\em Ansatz\/} has the required property and yields, as we shall see,
a few significant results:
\begin{equation}
\overline{\eta}_k(\pq)=\eta_k(p_k).
\label{eq:17}
\end{equation}
In that case, it is immediate to verify, using (\ref{eq:basic}), that (\ref{eq:16}) indeed holds. In this case,
the polynomial $\hhh(z|\pq)$ has the following explicit expression:
\begin{eqnarray}
\hhh(z|\pq)&=&\sum_{r=1}^N\eta_r(p_r)\prod_{k=1;(k\neq r)}^N\frac{z-q_k}{q_r-q_k}
\nonumber\\
&=&-\sum_{r=1}^N\frac{
\eta_r(p_r)}{
\prod_{k=1;(k\neq r)}^N(q_r-q_k)
}
\frac{\partial\sover(z|\qq)}{\partial q_r},
\label{eq:18}
\end{eqnarray}
where $\sover(z|\qq)$ is the (monic) polynomial defined in (\ref{eq:9}).
We now define a related polynomial $\eee(z|\qq)$ in the usual way:
\begin{subequations}
\begin{eqnarray}
\eee(q_k|\qq)&=&q_k^N,
\label{eq:elementary.a}
\\
\eee(z|\qq)&=&z^N-\sover(z|\qq):=\sum_{k=1}^Ne_k(\qq)z^{N-k}.
\label{eq:elementary.b}
\end{eqnarray}
\label{eq:elementary}
\end{subequations}
The $e_k(\qq)$ are therefore (up to an alternating sign) the elementary 
symmetric functions\cite{combin}
\begin{equation}
e_k(\qq)=(-1)^{k-1}\!\!\!
\!\!\!
\sum_{1\leq r_1<r_2<\cdots<r_k\leq N}\left(
\prod_{l=1}^kq_{r_l}
\right).
\label{eq:19}
\end{equation}
From (\ref{eq:18}) follows the following explicit expression for $h_k(\pq)$:
\begin{equation}
h_k(\pq)=\sum_{r=1}^N\eta_r(p_r)\frac{\partial e_k(\qq)}{\partial q_r}
\prod_{l=1;(l\neq r)}^N(q_r-q_l)^{-1}.
\label{eq:20}
\end{equation}
All such Hamilton\-ians are therefore integrable. Clearly, the Hamilton\-ian (\ref{eq:3})
belongs to this class, for $\eta_k(p)=e^p$ and for $k=1$ as do the simplest versions of
the Ruijsenaars--Schneider Hamilton\-ian introduced in \cite{RS}. We have now explicitly found the expression 
for the corresponding action variables which are simply given by $h_k(\pq)$ for 
$2\leq k\leq N$. 

\subsection{Separation of variables}
\label{subsec:sepvar}

Using the above technique, it is also straightforward to solve the Hamilton--Jacobi equation for 
one of the Hamilton\-ians of the class (\ref{eq:20}), or in fact, more generally, for all Hamilton\-ians 
satisfying (\ref{eq:16}). 

To this end, one makes the following observation: since all the $h_k(\pq)$ derived from
$\hhh(z|\pq)$ are constants of the motion for any of them, it follows that, for all the dynamics 
defined by some $h_k(\pq)$, one has
\begin{equation}
\hhh\left[
z|\underline{p}(t), \underline{q}(t)
\right]
=P_0(z),
\label{eq:21}
\end{equation}
where $P_0(z)$ is  the constant polynomial
determined by the initial values of the constants of motion $h_k$. Using (\ref{eq:15b}) yields,
by substituting $z$ by $q_k(t)$:
\begin{equation}
P_0[q_k(t)]=\eta_k[p_k(t)].
\label{eq:22}
\end{equation}
Let us now define $\phi_k(x)$ as the inverse function of $\eta_k(p)$, that is
\begin{equation}
\phi_k[\eta_k(p)]=p.
\label{eq:23}
\end{equation}
We will assume in the following that $\phi_k(x)$ is a uniquely defined function. If this is not the case,
special precautions must be taken in each case. If we now define $S(\qq)$ so that 
\begin{equation}
p_k=\frac{\partial S}{\partial q_k},
\label{eq:24}
\end{equation}
it follows from (\ref{eq:22}, \ref{eq:23}) that
\begin{equation}
\frac{\partial S}{\partial q_k}=\phi_k\left[
P_0(q_k)
\right].
\label{eq:25}
\end{equation}
This equation has the manifest solution
\begin{equation}
S(\qq)=\sum_{k=1}^N\int_0^{q_k}dy\,\phi_k\left[
P_0(y)
\right].
\label{eq:25bis}
\end{equation}
The coefficients of the polynomial $P_0(y)$ provide $N$ arbitrary constants, which are the initial values
of the constants of the motion $h_l$, for $1\leq l\leq N$, so that we have here
a general solution of the Hamilton--Jacobi equation. Let us now define $\hh$ as the vector $(h_1,\ldots,h_N)$.
From the general theory, we have that the partial derivative of $S(\qq,\hh)$ with respect to $h_l$
is also a constant of motion $\beta_l$:
\begin{equation}
\beta_l=\frac{\partial S(\qq,\hh)}{\partial h_l}=\sum_{k=1}^N\int_0^{q_k}\frac{dy\,y^{N-l}}{\eta_k^\prime(
\phi_k[P_0(y)])}.
\label{eq:26}
\end{equation}
From this the motion can be recovered, if we remember that, if the dynamics we consider is that
of $h_k$, then $\beta_k$ corresponds to the time.

This also provides a very elementary way to reduce the Hamilton\-ian equations of motion, which involve
the $2N$ variables $p_k$ and $q_k$ for $1\leq k\leq N$, to a system of $N$ first order ordinary differential 
equations involving the $q_k$ only. Indeed, Hamilton's equations yield for the $q_k$:
\begin{equation}
\dot{q}_k=\eta_k^\prime(p_k)\prod_{r=1;(r\neq k)}^N(q_k-q_r)^{-1}.
\label{eq:26.1}
\end{equation}
But according to (\ref{eq:22}) we may substitute $p_k$ by $\phi_k[P_0(q_k)]$. This leads to 
\begin{equation}
\dot{q}_k=\eta_k^\prime
\left(
\phi_k[P_0(q_k)]
\right)\prod_{r=1;(r\neq k)}^N(q_k-q_r)^{-1}.
\label{eq:26.2}
\end{equation}
This result applied to the goldfish case, $\eta(p)=e^p$, 
leads to a result obtained in a quite different way by Calogero\cite{polsum}, namely that the solutions
of the system of ODE's
\begin{equation}
\dot{q}_k=P_0(q_k)
\prod_{r=1;(r\neq k)}^N(q_k-q_r)^{-1}
\label{eq:26.3}
\end{equation}
for any given fixed polynomial $P_0(q)$ are also a solution of the goldfish equations (\ref{eq:3}).
\subsection{An elementary special case}
\label{subsec:elementary}

A very elementary example is that in which $\eta_k(p)=p$. Then we also have $\phi_k(x)=x$,
and hence, from (\ref{eq:26})
\begin{equation}
\beta_l=\frac{1}{N-l+1}\sum_{k=0}^N q_k^{N-l+1}.
\label{eq:27}
\end{equation}
We thus see that for this Hamilton\-ian, all moments of the $q_k$ remain constant, except for that 
corresponding to the Hamilton\-ian under study, which corresponds to the time. If we wish to 
know the behavior of $h_1(\pq)$, then we set
\begin{equation}
t=\frac1N\sum_{k=1}^Nq_k^N,
\label{eq:28}
\end{equation}
whereas all other moments are constant. As is readily seen, this means that, if one considers
the monic polynomial of which the $q_k$ are the zeros, then its coefficients are all constant
except the constant term, which grows proportionally to time. 

The equations of motion corresponding to this Hamilton\-ian, see (\ref{eq:20}) with $k=1$, are
\begin{equation}
\dot{q}_j=\prod_{k=1;(k\neq j)}^N(q_j-q_k)^{-1}.
\label{eq:29}
\end{equation}
The same result was found by a completely different approach as Exercise 2.3.4.1-12 in \cite{hamil1}.

\section{Translation invariance}
\label{sec:translation}

For the specific case mainly discussed above, in which the functions $\overline{\eta}(\pq)$ defined
in (\ref{eq:17}) only depend on the $N$-vector $\pp$, the issue of translation invariance has interesting
ramifications. It is easily seen that the Hamilton\-ian $h_1(\pq)$ is, in that case, always invariant under 
translations, so that, if we set
\begin{equation}
P=\sum_{k=1}^Np_k,
\label{eq:Pdef}
\end{equation}
one obtains
\begin{equation}
\poiss{h_1(\pq)}{P}=0.
\label{eq:30}
\end{equation}
This can be simply understood: since the values of $\hhh(q_k|\pq)$ do not depend on $\qq$, we
have, if we define $\aaaa$ as the $N$-vector $(a,\ldots,a)$
\begin{equation}
\hhh(z-a|\pq)=\hhh(z|\pp,\qq+\aaaa).
\label{eq:31}
\end{equation}
Upon translation of $z$, by a constant, 
the leading coefficient of $\hhh(z|\pq)$ is unchanged, from which the claim follows.
On the other hand, the other coefficients are, of course, affected, implying that the Hamilton\-ians
$h_k(\pq)$ for $2\leq k\leq N$ are not translation invariant, as indeed follows from the explicit
expression (\ref{eq:20}). However, by differentiating (\ref{eq:31}) with respect to $a$, one obtains
\begin{eqnarray}
\poiss{\hhh(z|\pp,\qq)}{P}&=&\frac{\partial\hhh(z|\pq)}{\partial z}
\nonumber\\
&=&\sum_{k=1}^N(N-k)
h_k(\pq)z^{N-k-1}.
\label{eq:32}
\end{eqnarray}
From this follows
\begin{equation}
\poiss{h_k(\pq)}{P}=(N-k+1)h_{k-1}(\pq),
\label{eq:33}
\end{equation}
and therefore, for any $1\leq k\leq N$, one has
\begin{equation}
\poiss{h_k(\pq)}{\poiss{h_k(\pq)}{P}}=0.
\label{eq:34}
\end{equation}
Thus, if we consider the dynamics determined by any given $h_k(\pq)$, we find
\begin{equation}
\ddot P=0. 
\label{eq:35}
\end{equation}
We therefore see that for all Hamilton\-ians, total momentum $P$ is either conserved or 
a measure of time. 
\section{Remarkable properties of the ``goldfish'' Hamilton\-ian}
\label{sec:goldfish}
The various Hamilton\-ians we have up to now considered are essentially of the type corresponding to free 
motion. They correspond to various systems having a type of modified kinetic energy. We do not have 
integrable cases involving Hamilton\-ians of the type kinetic plus potential energy, that is, we do not have the
possibility of adding purely $\qq$ dependent terms while maintaining integrability.

In the case of the ``goldfish'' Hamilton\-ian, it is possible to use the same approach to go considerably further.
As we shall see, it is possible to obtain results for a rather broad class of combinations between 
the $h_k(\pq)$ and the $e_k(\qq)$. 
This is due to the fact that the polynomial $\eee(z|\qq)$ has a Poisson bracket with the polynomial
$\hhh(z|\pq)$ which can in turn be expressed in terms of $\eee(z|\qq)$ and $\hhh(z|\pq)$. Let us show this
now, and then consider the consequences.

\subsection{Poisson brackets for the ``goldfish'' Hamilton\-ian}
\label{subsec:PB}
Let us define $\rrr(z,w|\pq)$ as 
\begin{equation}
\rrr(z,w|\pq)=\poiss{\hhh(w|\pq)}{\eee(z|\qq)},
\label{eq:35.1}
\end{equation}
where we are now limiting ourselves to the case of the goldfish Hamilton\-ian defined by 
$\eta_k(p)=e^p$. In Appendix \ref{app:b} we show, using (\ref{eq:basic}), that
\begin{equation}
\rrr(q_k,q_l|\pq)=e^{p_k}\delta_{k,l}\prod_{r=1;(r\neq k)}^N(q_k-q_r).
\label{eq:new1}
\end{equation}
This has several interesting consequences. First and foremost, it shows that the polynomial $\ccc(z,w|\pq)$
is {\em symmetric\/} in $z$ and $w$, thereby implying
\begin{equation}
\poiss{h_k(\pq)}{e_l(\qq)}=\poiss{h_l(\pq)}{e_k(\qq)}.
\label{eq:new1.01}
\end{equation}
Second, we find an explicit expression for $\rrr(z,w|\pq)$, namely
\begin{eqnarray}
\rrr(z,w|\pq)&=&\sum_{r=1}^Ne^{p_r}
\left(
\prod_{l=1;(l\neq r)}^N(z-q_l)
\right)\nonumber\\
&&\quad\times
\left(
\prod_{k=1;(k\neq r)}^N\frac{w-q_k}{q_r-q_k}
\right).
\label{eq:new1.02}
\end{eqnarray}
From (\ref{eq:new1.02}) readily follows that the leading term in $\rrr(z,w|\pq)$, that is, the polynomial 
in $w$ premultiplying $z^{N-1}$ is given by
\begin{equation}
\sum_{k=1}^N\poiss{h_1(\pq)}{e_k(\qq)}w^{N-k}=\hhh(w|\pq),
\label{eq:new1.03}
\end{equation}
since it satisfies the defining equation (\ref{eq:15b}).
This therefore implies that 
\begin{equation}
\poiss{h_1(\pq)}{e_k(\qq)}=\poiss{h_k(\pq)}{e_1(\qq)}=h_k(\pq).
\label{eq:new1.04}
\end{equation}

Finally, from (\ref{eq:new1}), we immediately observe that
\begin{equation}
(z-w)\rrr(z,w|\pq)\equiv0,
\label{eq:new2}
\end{equation}
where we define the equivalence relation as stating that the left-hand side vanishes for all
possible choices of $z$ and $w$ among the $q_k$. 

From this one obtains, after some calculations detailed in Appendix \ref{app:c}, an 
explicit expression for the Poisson brackets $\poiss{h_k(\pq)}{e_l(\qq)}$ given by
\begin{widetext}
\begin{eqnarray}
\poiss{h_k(\pq)}{e_{l}(\qq)}&=&
\sum_{m=l}^{k+l-2}\left[
e_{k+l-m-1}(\qq)h_m(\pq)
-
e_{m}(\qq)h_{k+l-m-1}(\pq)
\right]
+h_{k+l-1}(\pq)\nonumber\\
&=&\sum_{m=l}^{k+l-1}\left[
e_{k+l-m-1}(\qq)h_m(\pq)
-
e_{m}(\qq)h_{k+l-m-1}(\pq)
\right]
,
\label{eq:new3}
\end{eqnarray}
\end{widetext}
where, in the second form of the equation, we adopt the convention suggested in Appendix \ref{app:c},
that $e_0(\qq)=-1$ and $h_0(\pq)=0$. 
This relation, which expresses the Poisson brackets of the $h_k(\pq)$ and the 
$e_k(\qq)$ bilinearly in terms of the same variables, will play a basic role in the following. 
The relation's mathematical significance is somewhat unclear to me, as it clearly does 
not correspond to a Lie algebra, but it turns out to be quite powerful. 

Summarising, the Poisson bracket of the $e_k$ and the $h_l$ are characterised by 3 fundamental properties:
\begin{enumerate}
\item identity: 
\begin{equation}
\poiss{h_k(\pq)}{e_1(\qq)}=h_k(\pq).
\label{eq:new3.1}
\end{equation}

\item symmetry:
\begin{equation}
\poiss{h_k(\pq)}{e_l(\qq)}=
\poiss{h_l(\pq)}{e_k(\qq)}.
\label{eq:new3.2}
\end{equation}

\item bilinear closure: the Poisson brackets of any $e_k(\qq)$ with any $h_l(\pq)$ can be expressed
as a sum of products of $h_r(\pq)$ and $e_s(\qq)$.
\end{enumerate}

It is from these three properties alone that all the following results are obtained. 

\subsection{Explicit solution for all Hamilton\-ians $h_k(\pq)$}
\label{subsec:sol}

To solve for the dynamics of the Hamilton\-ian $h_k(\pq)$ for any given $k$, it is clearly sufficient to 
determine the time-dependence of the $e_j(\qq)$ for $1\leq j\leq N$. Indeed, the $h_j(\pq)$ are 
all constant. The $e_j(\qq)$ are then sufficient to determine the $q_l$ by the algebraic process of 
computing the roots of $\sover(z|\qq)$. 

The equations obeyed by the $e_j(\qq)$ are simply given by
\begin{equation}
\dot{e}_j(\qq)=\poiss{h_k(\pq)}{e_j(\qq)}.
\label{eq:new4}
\end{equation}
Using (\ref{eq:new3}) and the constancy of the $h_j(\pq)$ for all $j$, we obtain the following 
expression for the equation of motion
\begin{equation}
\dot{e}_j(\qq)=\sum_{l=1}^Na^{(k)}_{j,l}(h_1,\ldots,h_N)e_l(\qq)+b_j^{(k)}(h_1,\ldots,h_N).
\label{eq:new5}
\end{equation}
This can be rewritten more simply using the convention, introduced in Appendix \ref{app:c}, that $e_0(\qq)=-1$.
One then has
\begin{equation}
\dot{e}_j(\qq)=\sum_{l=0}^Na^{(k)}_{j,l}(h_1,\ldots,h_N)e_l(\qq).
\label{eq:new5.1}
\end{equation}
Here the $a^{(k)}_{j,0}$ have replaced the $b^{(k)}_j$.
Here the $a^{(k)}_{j,l}$ and the $b_j^{(k)}$ are linear expressions in the $h_j$, where 
the $h_j$ denote the initial values which the
integrals of motion $h_j(\pq)$ take at the beginning of the evolution. 

Some additional notation will allow to formulate this result more compactly: denote by $\ee(\qq)$ the 
vector $(e_0(\qq),e_1(\qq),\ldots,e_N(\qq))$ and by $\hh$ the vector $(h_0(\pq), h_1,\ldots,h_N)$. There then exists
a matrix ${\bf A}^{(k)}$ such that 
\begin{equation}
\dot{\ee}={\bf A}^{(k)}(\hh)\ee.
\label{eq:new6}
\end{equation}

The $e_j(\qq)$ therefore undergo a linear time evolution, the matrix of which depends on the $h_j$. 
However, since these are constant, they may be viewed as parameters. The matrix ${\bf A}^{(k)}$ 
additionally depends on $k$, corresponding to the specific Hamilton\-ian $h_k(\pq)$ we are 
considering. Since, however, the flows of the various $h_k(\pq)$ commute, it follows that for 
all $\hh$
\begin{equation}
\left[
{\bf A}^{(k)}(\hh),{\bf A}^{(l)}(\hh)
\right]=0,
\label{eq:new7}
\end{equation}
leading to the interesting consequence that the ${\bf A}^{(k)}$ have eigenvectors that can be 
chosen independent of $k$. Note further that one can find an explicit expression for ${\bf A}^{(k)}$
using the fundamental relation (\ref{eq:new3}). 

Finally, it is also quite easy to rederive the solution of the original ``goldfish'' Hamilton\-ian, $h_1(\pq)$.
In this case we have
\begin{equation}
\dot{e}_k(\qq)=\poiss{h_1(\pq)}{e_k(\qq)}=h_k(\pq).
\label{eq:new8}
\end{equation}
Since $h_k$ is conserved, one verifies immediately that the $e_k(\qq)$ move at a constant velocity, 
a result which has been derived in many different ways earlier\cite{goldfish, hamil1}. 

\subsection{Integrals of motion for all Hamilton\-ians}
\label{subsec:int-mot}

From the formalism developed above, it is straightforward to derive $2N-1$ integrals of motion for 
an arbitrary  Hamilton\-ian $h_k(\pq)$. First, we have all the $h_j(\pq)$ for $1\leq j\leq N$. To obtain the
other $N-1$ integral of motion, we use the solution given by (\ref{eq:new6}) in subsection \ref{subsec:sol}.
Using this, we can write the initial conditions $\ee(0)$ as a function of the time $t$ and the $\ee(t)$ as follows:
\begin{equation}
\ee(0)=\exp\left(
-t{\bf A}^{(k)}
\right)\ee(t).
\label{eq:new9}
\end{equation}
Since the $\ee(0)$ are clearly conserved quantities, we have a set of $N$ time-dependent conserved
quantities. We require a way to eliminate the time-dependence. This is readily done using the observation
made in Section \ref{sec:translation}: the total momentum $P$, defined in (\ref{eq:Pdef}), is proportional to the time.
To be precise, one has from (\ref{eq:33}) that the quantity
\begin{equation}
\tau=\frac{P}{(N-k+1)h_{k-1}}
\label{eq:new10}
\end{equation}
differs from $t$ by an additive constant, so that 
\begin{equation}
\underline{\Phi}(\pq)=\exp\left[
-\tau(\pq){\bf A}^{(k)}(\hh(\pq))
\right]\ee(\qq)
\label{eq:new11}
\end{equation}
is a vector of time-independent conserved quantities of $h_k(\pq)$. They are never functionally
independent, but exactly one can always be expressed in terms of all others, thereby leading 
to the desired result. 

The above technique fails in the case of $h_1(\pq)$, since $P$ is then a conserved quantity. In that 
case, a different approach leads more rapidly to an equivalent result: the quantities
\begin{equation}
\Lambda_{k,l}=e_k(\qq)h_l(\pq)-e_l(\qq)h_k(\pq)
\label{eq:new12}
\end{equation}
all commute with $h_1(\pq)$, as is readily checked using property (1) of the Poisson bracket, as described in 
subsection \ref{subsec:PB}. These are functionally independent of the $h_j(\pq)$ and the functions 
$\Lambda_{1,k}$, for $2\leq k\leq N$ are also functionally independent
from one another, so that we have found $2N-1$ integrals of motion. 

\subsection{Two other sets of exactly solvable Hamilton\-ians}
\label{subsec:sol-more}

The following set of Hamilton\-ians
\begin{equation}
\htilde_k(\pq)=h_k(\pq)+\alpha e_k(\pq)
\label{eq:new13}
\end{equation}
all commute with each other, as straightforwardly follows from the symmetry property
of the Poisson bracket described in subsection \ref{subsec:PB}. All these Hamilton\-ians
are therefore {\em integrable\/} in the sense of Liouville. Note that this result can also 
be obtained directly from (\ref{eq:basic}) by showing that the polynomial $\tilde{\hhh}(z|\pq)$
defined by
\begin{equation}
\tilde{\hhh}(q_k|\pq)=e^{p_k}+\alpha q_k^N
\end{equation}
indeed defines a Poisson-commuting family, that is, that it satisfies (\ref{eq:16}). This shows that 
there is a large variety of possible solutions of (\ref{eq:16}) beyond the range of the family defined 
in (\ref{eq:20}).

Let us now first consider $\htilde_1(\pq)$. Addditionally to the integrals of motion $\htilde_k(\pq)$,
we find also $h_k(\pq)/h_l(\pq)$. The integration then becomes entirely straightforward: 
$h_1$ satisfies the equation
\begin{equation}
\dot{h}_1(\pq)=-\alpha h_1(pq).
\label{eq:new13.1}
\end{equation}
If we denote by $\rho_k$ the constant value $h_k/h_1$, then the $h_k$ are determined
as $\rho_k h_1$, whereas, if $\htilde_k$ denotes the constant value of $\htilde_k(\pq)$,
the $e_k(\qq)$ can be obtained directly from the $h_k$. 
Note further that, if $\alpha$ is a pure imaginary number $i\omega$, then $\htilde_1(\pq)$ is an 
{\em isochronous\/} Hamilton\-ian, since in that case, $h_1$ is periodic with period
$2\pi/\omega$. The $e_k(\qq)$ and the $h_l(\pq)$ therefore execute a periodic motion,
implying that the $q_k$ do so as well, though possibly with a larger period, see \cite{matteo}. 

However, the equations of motion for the Hamilton\-ians $\htilde_k(\pq)$ 
for $2\leq k\leq N$, cannot be integrated 
in the same way, nor have I been able to identify for these a full set of $2N-1$
set of integrals of motion. 

On the other hand the more general Hamilton\-ian
\begin{equation}
H(\pq)=\sum_{k=1}^N \lambda_kh_k(\pq)+\mu e_1(\qq)
\label{eq:new14}
\end{equation}
can be solved explicitly: once more, the quantities $h_k(\pq)/h_l(\pq)$ are constants of motion,
and further
\begin{equation}
\dot{h}_1(\pq)=-\mu h_1(\pq),
\label{eq:new15}
\end{equation}
so that we can compute explicitly all $h_k$. The equations of motion for the $e_k$ are
\begin{equation}
\dot{e}_k(\qq)=\sum_{l,m=1}^N c^{(k)}_{l,m}e_l(\qq)h_m(\pq),
\label{eq:new16}
\end{equation}
where the $c^{(k)}_{l,m}$ are constants, as follows from the closure property. If we now change the time variable
to $\tau$ defined by
\begin{equation}
d\tau=h_1(t)dt=h_1(0)e^{-\mu t},
\label{eq:new17}
\end{equation}
one obtains
\begin{equation}
\frac{d{e}_k(\qq)}{d\tau}=\sum_{l,m=1}^N c^{(k)}_{l,m}\rho_m e_l(\qq).
\label{eq:new18}
\end{equation}
As a function of $\tau$, the $e_k$ are thus the solutions of a linear system of ordinary differential equations,
which can be solved in an elementary way. Again, if $\mu$ is a purely imaginary number $i\omega$,
the system is isochronous with a period of $2\pi/\omega$. 


\section{Conclusions}
\label{sec:conclusions}

Summarising, we present an elementary approach to obtaining sets
of integrable Hamilton\-ians.
Our main result can be summarised as saying that all Hamilton\-ians of the form stated in 
(\ref{eq:20}) are integrable. The results of Subsection \ref{subsec:sepvar} show that 
one can also reduce the problem explicitly to quadratures. This class of Hamilton\-ians
includes, in particular, some forms of the Ruijse\-naars--Schneider\cite{RS} Hamilton\-ian
as well as the goldfish Hamilton\-ian introduced in \cite{hamil}. For the latter, as shown 
in Section \ref{sec:goldfish}, a large number of additional results can be derived, such as
an explicit description of the $2N-1$ constants of motion corresponding to the maximally 
superintegrable nature of the dynamics. 
\begin{acknowledgments}
I wish to acknowledge frequent conversations with Francesco Calogero, as well as with 
A.V.~Mikhailov and A.~Pogrebkov in the framework of the Meeting on Integrable and Quasi-integrable
Systems organized by the Centro Internacional de Ciencias from November 14th to December 9th
2016, as well as UNAM--DGAPA--PAPIIT IN103017 and CONACyT 254515 for financial support. 
\end{acknowledgments}

\appendix

\section{Calculations leading to (\ref{eq:basic})}
\label{app:a}

Starting from (\ref{eq:12}), it is readily seen that
\begin{eqnarray}
&&\ccc(q_k,q_l|\pq)=\sum_{r,s=1}^N
\bigg[
\poiss{a_r(\pq)q_k^{N-r}}{b_s(\pq)q_l^{N-s}}\nonumber\\
&&\qquad-\poiss{q_k^{N-r}}{b_s(\pq)q_l^{N-s}}a_r(\pq)\nonumber\\
&&\qquad-\poiss{a_r(\pq)q_k^{N-r}}{q_l^{N-s}}b_s(\pq)
\bigg].
\label{eq:13}
\end{eqnarray}
Here the first term arises from the definitions of $\alpha_k(\pq)$ and $\beta_l(\pq)$ as given in 
(\ref{eq:7b}) and (\ref{eq:10b}) respectively. The second term is further evaluated by noting that 
\begin{equation}
\poiss{q_k^{N-r}}{b_s(\pq)q_l^{N-s}}=-(N-r)q_k^{N-r-1}q_l^{N-s}\frac{\partial b_s(\pq)}{\partial q_k}
\end{equation}
with a similar result for the third term. We therefore finally find:
\begin{eqnarray}
&&\ccc(q_k,q_l|\pq)=\poiss{\alpha_k(\pq)}{\beta_l(\pq)}\nonumber\\
&&\qquad+\sum_{r,s=1}^N\bigg[
(N-r)a_r(\pq)q_k^{N-r-1}
\frac{\partial b_s(\pq)}{\partial p_k}q_l^{N-s}
\nonumber\\
&&\qquad-(N-s)b_s(\pq)q_l^{N-s-1}
\frac{\partial a_r(\pq)}{\partial p_l}q_k^{N-r}
\bigg],
\label{eq:14}
\end{eqnarray}
which readily yields the final result (\ref{eq:basic}). 

\section{Computation of the Poisson bracket  (\ref{eq:new1}) of $\eee(z\qq)$ with $\hhh(w|\pq)$ for the goldfish Hamilton\-ian}
\label{app:b}

In the following we apply formula (\ref{eq:basic}) to the polynomials $\hhh(w|\pq)$ and   $\eee(z|\qq)$, 
in this order. The first term is given by 
\begin{equation}
\poiss{e^{p_l}}{q_k^N}=Nq_k^{N-1}e^{p_l}\delta_{k,l}.
\label{eq:b1}
\end{equation}
The second term vanishes, since 
\begin{equation}
\left.\frac{\partial\hhh(z|\pq)}{\partial z}\right|_{z=q_l}\frac{\partial e_k(\qq)}{\partial p_l}=0,
\label{eq:b3}
\end{equation}
and the third term is
\begin{eqnarray}
\left.\frac{\partial\eee(w|\qq)}{\partial w}\right|_{w=q_k}e^{p_l}\delta_{k,l}&=&\left(
Nq_k^{N-1}-
\prod_{r=1;(r\neq k)}^N
(q_k-q_r)
\right)\nonumber\\
&&\quad\times
e^{p_k}\delta_{k,l}.
\label{eq:b2}
\end{eqnarray}
Summing all these contributions with the appropriate signs leads to the desired result (\ref{eq:new1}). 

\section{Computation of the recursion relations for the Poisson brackets of $e_k(\qq)$ and $h_l(\pq)$}
\label{app:c}

Starting from (\ref{eq:new2}), we cannot state that the polynomial on the left-hand side is zero, since
it has degree $N$ in both $z$ and $w$, so that it is not uniquely determined by its values on 
all $N$ values $\qq$. But we can say that
\begin{eqnarray}
&&(z-w)\rrr(z,w|\pq)=z^N\hhh(w|\pq)-w^N\hhh(z|\pq)\nonumber\\
&&\qquad+\sum_{k=2}^{N-1}\sum_{l=1}^{N-1}
\poiss{h_k(\pq)}{e_l(\qq)}
z^{N-k+1}w^{N-l}
\nonumber\\
&&\qquad-
\sum_{k=1}^{N-1}\sum_{l=2}^{N-1}
\poiss{h_k(\pq)}{e_l(\qq)}
z^{N-k}w^{N-l+1}
.
\label{eq:c1}
\end{eqnarray}
Note that we have here separated explicitly those terms in which $z$ or $w$
appears with the order $N$, from all other terms in which they all appear with lesser order. 
On the other hand, we have
\begin{equation}
z^N\equiv\eee(z|\qq),
\label{eq:c2}
\end{equation}
since $\sover(z|\qq)\equiv0$. We thus obtain from (\ref{eq:c1}).
\begin{eqnarray}
&&\eee(z|\qq)\hhh(w|\pq)-\eee(w|\qq)\hhh(z|\pq)\nonumber\\
&&\qquad+\sum_{k=1}^{N-2}\sum_{l=1}^{N-1}
\poiss{h_{k+1}(\pq)}{e_l(\qq)}
z^{N-k}w^{N-l}
\nonumber\\
&&\qquad-
\sum_{k=1}^{N-1}\sum_{l=1}^{N-2}
\poiss{h_k(\pq)}{e_{l+1}(\qq)}
z^{N-k}w^{N-l}
\nonumber\\
&&\qquad=0.
\label{eq:c3}
\end{eqnarray}
Here we use the equality sign instead of equivalence, since the polynomial on the
left-hand side has degree $N-1$. 

From this now follows by identifying terms, that 
\begin{widetext}
\begin{equation}
e_k(\qq)h_l(\pq)-e_l(\qq)h_k(\pq)+
\poiss{h_{k+1}(\pq)}{e_l(\qq)}-
\poiss{h_k(\pq)}{e_{l+1}(\qq)}
=0.
\label{eq:c4}
\end{equation}
\end{widetext}
This recursion allows to express any Poisson bracket between an $e_k(\qq)$ and an $h_l(\pq)$
as a sum of products of $e_r(\qq)$ and $h_s(\pq)$. More explicitly we have
\begin{eqnarray}
&&\poiss{h_k(\pq)}{e_{l+1}(\qq)}=
\sum_{m=l}^{k+l-1}\bigg[
e_{k+l-m}(\qq)h_m(\pq)
\nonumber\\
&&\qquad-
e_{m}(\qq)h_{k+l-m}(\pq)
\bigg]
+h_{k+l}(\pq).
\label{eq:c5}
\end{eqnarray}
Note finally that the recursion (\ref{eq:c4}) in its present form only holds for $1\leq k,l\leq N-1$. When, for
example, $k=0$, the expression $\poiss{h_1}{e_l}$ arises, which is equal to $h_l$ and thus not quadratic
in the $\ee$ and $\hh$. To extend the recursion appropriately, it is sufficient to choose $e_0(\qq)=-1$ and 
$h_0(\pq)=0$. With these values the recursion extends to the range $0\leq k,l\leq N-1$ and allows 
to express any Poisson bracket of the $\ee$ with the $\hh$ as a sum of products of $e_k$ with $h_l$, 
with $0\leq k,l\leq N$.



\begin{thebibliography}{99}
\bibitem{calorig} F. Calogero ``The neatest many-body problem amenable to
exact treatments (a ``goldfish''?)'' 
Physica D {\bf152--153} 78--84 (2001) 


\bibitem{goldfish} F. Calogero ``Motion of poles and zeros of special solutions 
of nonlinear and linear partial differential equations, and related ÔsolvableÕ 
many-body problems'' Nuovo Cimento {\bf43B}, 177--241 (1978) 

\bibitem{hamil1} F. Calogero. {\em Classical many-body problems amenable to exact treatments},
Lecture Notes in Physics, Springer, Berlin (2001).

\bibitem{hamil} F. Calogero and J.-P. Fran\c coise, 
``Hamilton\-ian character of the motion of the zeros of a
polynomial whose coefficients oscillate over time'',
J. Phys. A: Math. Gen. {\bf30}, 211--218 (1997)

\bibitem{periodic} F. Calogero, ``A class of integrable Hamilton\-ian systems 
whose solutions are (perhaps) all completely periodic'', J. Math. Phys. {\bf38},  5711 (1997)

\bibitem{GF-QM1} M.C. Nucci, ``Quantizing preserving Noether symmetries'', J. of Nonlinear
Mathematical Physics {\bf 20} (3) 451--463 (2013)

\bibitem{GF-QM2} M.C. Nucci, ``Calogero's ÒgoldfishÓ is indeed a school of free particles''
J. Phys. A: Math. Gen. {\bf37}, 11391--11400 (2004)

\bibitem{morosi1} C. Morosi and G. Tondo, ``Quasi-bi-Hamilton\-ian systems and separability''
J. Phys. A {\bf30},  2799--2806 (1997)

\bibitem{blaszak}M. B\l{}aszak, ``On separability of bi-Hamilton\-ian chain with degenerated Poisson structures'' 
J. Math. Phys. {\bf39}, 3213--3235 (1998)

\bibitem{morosi2} G. Tondo and C. Morosi, ``Bi-Hamilton\-ian manifolds, quasi-bi-Hamilton\-ian 
systems and separation variables'',
Rep. Math. Phys. {\bf44} (1999), 255--266 (1999)

\bibitem{tondo} G. Tondo and P. Tempesta, ``Haantjes Structures for the Jacobi--Calogero Model
and the Benenti Systems'',
Symmetry, Integrability and Geometry: Methods and Applications SIGMA {\bf12}, 023 (2016)

\bibitem{combin}I.G. Macdonald, {\em Symmetric Functions and Hall Polynomials}, 
second ed. Oxford: Clarendon Press. (1995)

\bibitem{RS} S.N.M. Ruijsenaars and H. Schneider, ``A new class of integrable systems and 
its relation to solitons'', Ann. Phys. {\bf170}, 370--405 (1986)

\bibitem{polsum}  F. Calogero, ``Properties of the Zeros of the Sum
of two Polynomials'', J. Nonlin. Math. Physics 348--354 (2013)

\bibitem{matteo} D. Gomez-Ullate and M. Sommacal,  ``Periods of the goldfish 
many-body problem'', Journal of Nonlinear Mathematical Physics, 
{\bf12}:sup1, 351--362 (2005)





\end{thebibliography}
\end{document}